\documentclass[prd,aps,showkeys,floats,onecolumn]{revtex4}

\usepackage{amsmath}
\usepackage{amssymb}

\def\FFr{\displaystyle \frac}
\def\IIn{\displaystyle \int}
\def\S{\displaystyle \sum}

\begin{document}

\title{Classical analog of extended phase space SUSY and its breaking}
\author{G Ter-Kazarian}
 \email{gago_50@yahoo.com}

\address{Byurakan Astrophysical Observatory,
Byurakan 378433, Aragatsotn District, Armenia}

\begin{abstract}
We derive the classical analog of the extended phase space quantum
mechanics of the particle with odd degrees of freedom which gives
rise to (N=2)-realization of supersymmetry (SUSY) algebra. By means
of an iterative procedure, we find the approximate groundstate
solutions to the extended Schr\"{o}dinger-like equation and use
these solutions further to calculate the parameters which measure
the breaking of extended SUSY such as the groundstate energy.
Consequently, we calculate a more practical measure for the SUSY
breaking which is the expectation value of an auxiliary field. We
analyze non-perturbative mechanism for extended phase space SUSY
breaking in the instanton picture and show that this has resulted
from tunneling between the classical vacua of the theory. Particular
attention is given to the algebraic properties of shape invariance
and spectrum generating algebra.
\end{abstract}


\maketitle

\section{Introduction}
An interesting question of keeping the symmetry between canonical
coordinates and momenta in the process of quantization deserves an
investigation. From its historical development, this aspect of
statistical quantum mechanics, unfortunately, has attracted little
attention. However, much use has been made of the technique of
ordering of canonical coordinates (q) and momenta (p) in quantum
mechanics~\cite{rf:1,rf:11}. It was observed that the concept of an
{\it extended} Lagrangian, ${\cal L}(p,\,q,\,\dot{p},\,\dot{q})$  in
phase space allows a subsequent extension of Hamilton's principle to
actions minimum along the actual trajectories in $(p,\,q)-$, rather
than in $q-$space. This leads to the phase space formulation of
quantum mechanics. Consequently this formalism was developed further
in~\cite{rf:2} by addressing the extended phase space stochastic
quantization of Hamiltonian systems with first class holonomic
constraints. This in a natural way results in the Faddeev-Popov
conventional path-integral measure for gauge systems. Continuing
along this line in the present article we address the classical
analog of the extended phase space (N=2)-SUSY quantum
mechanics~~\cite{rf:10} of the particles which have both bosonic and
fermionic degrees of freedom, i.e., the quantum field theory in
$(0+1)$-dimensions in $(q,p)-$space, exhibiting supersymmetry (for
conventional SUSY quantum mechanics see~\cite{rf:3}-\cite{rf:72}).
We analyze in detail the non-perturbative mechanism for
supersymmetry breaking in the instanton picture~(\cite{rf:8}). This
paper has been organized as follows. In the first part (Sects. 2,
3), we derive the classical analog of the extended phase space SUSY
quantum mechanics and obtain the integrals of motion. Consequently,
we describe the extended phase space (N=2)-SUSY algebra. In the
second part (Sects. 4 - 5), by means of an iterative scheme, first,
we find the approximate groundstate solutions to the extended
Schr\"{o}dinger-like equation, and then calculate the parameters
which measure the breaking of extended SUSY such as the groundstate
energy. We calculate a more practical measure for the SUSY breaking,
in particular in field theories which is the expectation value of an
auxiliary field.  We analyze non-perturbative mechanism for extended
phase space SUSY breaking in the instanton picture and show that
this has resulted from tunneling between the classical vacua of the
theory. The section 6 deals with the independent group theoretical
methods with nonlinear extensions of Lie algebras from the
perspective of extended phase space SUSY quantum mechanics and,
further, shows how it can be useful for spectrum generating algebra.
The concluding remarks are given in section 7. Unless otherwise
stated we take the geometrized units ($\hbar=c=1$). Also, an
implicit summation on repeated indices are assumed throughout this
paper.

\section{The integrals of motion}
For the benefit of those not familiar with the framework of extended
phase space quantization, enough details are given below to make the
rest of the paper understandable. The interested reader is invited
to consult the original papers~\cite{rf:1,rf:11} for further
details. In the framework of the proposed formalism, the extended
Lagrangian can be written as
\begin{equation}
{\cal L}_{ext}(p,\,q,\,\dot{p},\,\dot{q})=-\dot{q}_{i}\,p_{i}
-q_{i}\,\dot{p}_{i} +{\cal L}^{q}+{\cal L}^{p}, \label{R1}
\end{equation}
where a dynamical system with N degrees of freedom described by the
2N independent coordinates $q=(q_{1},\ldots\, q_{N})$ and momenta
$p=(p_{1},\ldots\, p_{N})$ which are not, in general, canonical
pairs. A Lagrangian ${\cal L}^{q}(q,\,\dot{q})$ is given in
$q-$representation and the corresponding ${\cal L}^{p}(p,\,\dot{p})$
in $p-$representation.. A dot will indicate differentiation with
respect to $t$. The independent nature of $p$ and $q$ gives the
freedom of introducing a second set of canonical momenta for both
$p$ and $q$ through the extended Lagrangian: $
\pi_{q_{i}}=\FFr{\partial {\cal L}_{ext}}{\partial \dot{q}_{i}}=
\FFr{\partial {\cal L}^{q}}{\partial \dot{q}_{i}}-p_{i}, \quad
\pi_{p_{i}}=\FFr{\partial {\cal L}_{ext}}{\partial \dot{p}_{i}}=
\FFr{\partial {\cal L}^{p}}{\partial \dot{p}_{i}}-q_{i}. \label{R2}
$ One may now defines an extended Hamiltonian
\begin{equation}
\begin{array}{l}
H_{ext}(p,\,q,\,\pi_{p},\,\pi_{q})= \pi_{q_{i}}\,\dot{q}_{i}+
\pi_{p_{i}}\,\dot{p_{i}}- {\cal L}_{ext}(p,\,q,\,\dot{p},\,\dot{q})=
H(p+\pi_{q}, \,q)-H(p,\, q+\pi_{p}),
\end{array}
\label{R3}
\end{equation}
where $H(p,\,q)=p_{i}\dot{q}_{i}- {\cal L
}^{q}=q_{i}\dot{p}_{i}-{\cal L}^{p}$ is the conventional Hamiltonian
of the system. In particular, vanishing of $\pi_{q}$ /or $\pi_{p}$
is the condition for $p$ and $q$ to constitute a canonical pair. In
the language of statistical quantum mechanics this choice picks up a
pure state ({\it actual path}). Otherwise, one is dealing with a
mixed state ({\it virtual path}).  One may, however, envisage that
the full machinery of the conventional quantum mechanical dynamics
is extendible to the extended dynamics as alluded to above.  Here
$p$ and $q$ will be considered as independent c-number operators on
the integrable complex function $\chi(q,p)$. One of the key
assumptions of extended phase space quantization~\cite{rf:1,rf:11}
is the differential operators and commutation brackets for $\pi_{p}$
and $\pi_{q}$ borrowed from the conventional quantum mechanics:
\begin{equation}
\begin{array}{l}
\pi_{q_{i}} =-i\FFr{\partial}{\partial\,q_{i}},\quad [\pi_{q_{i}},
q_{j}]=-i\delta_{ij},\quad \pi_{p_{i}}
=-i\FFr{\partial}{\partial\,p_{i}}, \quad [\pi_{p_{i}},
p_{j}]=-i\delta_{ij}.
\end{array}
\label{R4}
\end{equation}
Note also the following
\begin{equation}
\begin{array}{l}
\left[p_{i},q_{j}\right]=\left[p_{i},p_{j}\right]=\left[q_{i},q_{j}\right]=
\left[\pi_{p_{i}}, \pi_{q_{j}}\right]=
\left[\pi_{p_{i}},\pi_{p_{j}}\right]=
\left[\pi_{q_{i}},\pi_{q_{j}}\right]=0.
\end{array}
\label{R5}
\end{equation}
By the virtue of Eq.~(\ref{R4}) and Eq.~(\ref{R5}), $H_{ext}$ is now
an operator on $\chi$. Along the trajectories in $(p,\, q)$ space,
however, it produces the state functions, $\chi(p,\, q,\, t),$ via
the following Schr\"{o}dinger-like equation:
\begin{equation}
\begin{array}{l}
i\,\FFr{\partial}{\partial\,t}\chi=H_{ext}\,\chi.
\end{array}
\label{R6}
\end{equation}
Solutions of Eq.~(\ref{R6}) are
\begin{equation}
\begin{array}{l}
\chi(q,p,t)=\overline{\chi}_{r}(q,p,t)e^{-ipq}=a_{\alpha\beta}\,\psi_{\alpha}(q,\,
t)\,\phi_{\beta}^{*}(p,\, t)\,e^{-i\,pq},
\end{array}
\label{R113}
\end{equation}
where $ a=a^{\dag},\quad \mbox{positive definite}, \quad tr\, a=1,$
  and $\psi_{\alpha}$ and
$\phi_{\alpha}^{*}$ are solutions of the conventional
Schr\"{o}dinger equation in $q-$ and $p-$representations,
respectively. They are mutually Fourier transforms. Note that the
$\alpha$ and $\beta$ are not, in general, eigenindices. The
normalizable $\chi$  ($\int\,\chi\,dp\,dq=tr(a)=1$) is a physically
acceptable solution. The exponential factor is a consequence of the
total time derivative, $-d(qp)/dt,$ in Eq.~(\ref{R1}) which can be
eliminated. Actually, it is easily verified that
\begin{equation}
\begin{array}{l}
(p+\pi_{q})\chi_{r}(q,p,t)=(\pi_{q}\overline{\chi}_{r}(q,p,t))\,e^{-ipq},
\end{array}
\label {R29}
\end{equation}
and so on. Substitution of Eq.~(\ref{R29}) in Eq.~(\ref{R6}) gives
\begin{equation}
\begin{array}{l}
i\,\FFr{\partial}{\partial\,t}\,\overline{\chi}(q,p,t)=
\overline{H}_{ext}\,\overline{\chi}_{r}(q,p,t),
\end{array}
\label {R30}
\end{equation}
provided by the reduced Hamiltonian, $\overline{H}_{ext}$. From now
on we replace $H_{ext}$ by $\overline{H}_{ext}$, and
$\chi_{r}(q,p,t)$ by $\overline{\chi}_{r}(q,p,t)$, respectively, and
retain former notational conventions.

It is certainly desirable to derive the classical analog of the
extended phase space quantum mechanics of the particle with odd
degrees of freedom directly from what may be taken as the first
principle. Therefore, following~\cite{rf:7, rf:71, rf:72}, let us
consider a nonrelativistic particle of unit mass with two
($\alpha=1,2$) odd (Grassmann) degrees of freedom. The classical
extended Lagrangian Eq.~(\ref{R1}) can be written
\begin{equation}
\begin{array}{l}
{\cal L}_{ext}(p,\,q,\,\dot{p},\,\dot{q})=-\dot{q}\,p -q\,\dot{p}
+\FFr{1}{2}\,\dot{q}^{2}-F(q)+
\FFr{1}{2}\,\dot{p}^{2}-G(p)-R(q,\,p)\,N
+\FFr{1}{2}\,\psi_{\alpha}\dot{\psi}_{\alpha},
\end{array}
\label{R217}
\end{equation}
provided by $N=\psi_{1}\,\psi_{2}=-i\,\psi_{+}\,\psi_{-}.$ Here
$F(q):{\cal R}\rightarrow{\cal R},\quad$ $G(p):{\cal
R}\rightarrow{\cal R}$ and $R(q,\,p):{\cal R}\rightarrow{\cal R}$
are arbitrary piecewise continuously differentiable functions given
over the $1$-dimensional Euclidean space ${\cal R}$. The
$\psi_{\alpha}$ are two odd (Grassmann) degrees of freedom. The
nontrivial Poisson-Dirac brackets of the system Eq.~(\ref{R217}) are
\begin{equation}
\begin{array}{l}
\left\{q,\,\pi_{q}\right\}= 1,\quad \left\{p,\,\pi_{p}\right\}=
1,\quad
\left\{\psi_{\alpha},\,\psi_{\beta}\right\}=\delta_{\alpha\beta},\quad
\left\{\psi_{+},\, \psi_{-}\right\}=1,\quad \psi_{\pm}^{2}=0,\quad
\psi_{\pm}=\FFr{1}{\sqrt{2}}\,\left(\psi_{1}\pm i\psi_{2}\right).
\end{array}
\label {R219}
\end{equation}
The extended Hamiltonian $H_{ext}$ Eq.~(\ref{R3}) reads
\begin{equation}
\begin{array}{l}
H_{ext}(p,\,q,\,\pi_{p},\,\pi_{q})=
\FFr{1}{2}\,\left(p+\pi_{q}\right)^{2}+F^{2}(q)-
\FFr{1}{2}\,\left(q+\pi_{p}\right)^{2}-G^{2}(p)+ R(q,p)\,N,
\end{array}
\label{R222}
\end{equation}
which, according to Eq.~(\ref{R29}), reduces to
\begin{equation}
\begin{array}{l}
H_{ext}(p,\,q,\,\pi_{p},\,\pi_{q})=
\FFr{1}{2}\,\pi_{q}^{2}+F^{2}(q)- \FFr{1}{2}\,\pi_{p}^{2}-G^{2}(p)+
R(q,p)\,N.
\end{array}
\label{R322}
\end{equation}
The Hamiltonian Eq.~(\ref{R322}) yields the following equations of
motion:
\begin{equation}
\begin{array}{l}
\dot{q}=\pi_{q}, \quad \dot{p}=\pi_{p},\quad
\dot{\pi}_{q}=-F'_{q}(q)-R'_{q}(q,\,p)\,N,\quad
\dot{\pi}_{p}=-G'_{p}(p)+R'_{p}(q,\,p)\,N,\quad \dot{\psi}_{\pm}=\pm
iR(q,\,p)\,\psi_{\pm}.
\end{array}
\label{R223}
\end{equation}
A prime will indicate differentiation with respect either to $q$ or
$p$. Thus, $N$ is the integral of motion additional to $H_{ext}.$
Along the trajectories $q(t)$ and $p(t)$ in $(p,q)-$ spaces, the
solution to equations of motion for odd variables is
\begin{equation}
\begin{array}{l}
\psi_{\pm}(t)=\psi_{\pm}(t_{0})\,\exp \left[\pm
i\,\IIn^{t}_{t_{0}}\,R(q(\tau),\,p(\tau))\,d\tau\right].
\end{array}
\label{R224}
\end{equation}
Hence the odd quantities
\begin{equation}
\begin{array}{l}
\theta_{\pm}=\theta_{\pm}(t)\,\exp \left[\mp
i\,\IIn^{t}_{t_{0}}\,R(q(\tau),\,p(\tau))\,d\tau\right]
\end{array}
\label{R224}
\end{equation}
are nonlocal in time integrals of motion. In trivial case $R=0,$ we
have $\dot{\psi}_{\pm}=0,$ and $\theta_{\pm}=\theta_{\pm}.$ Suppose
the system has even complex conjugate quantities $B_{q,p\pm},\quad
(B_{q,p+})^{*}=B_{q,p-},$ whose evolution looks up to the term
proportional to $N$ like the evolution of odd variables in
Eq.~(\ref{R223}). Then local odd integrals of motion could be
constructed in the form
\begin{equation}
\begin{array}{l}
Q_{q,p\pm}=B_{q,p\mp}\,\psi_{\pm}.
\end{array}
\label{R225}
\end{equation}
Let us  introduce the oscillator-like bosonic variables $B_{q,p\pm}$
in $q-$ and $p$-representations
\begin{equation}
\begin{array}{l}
B_{q\mp}:{\cal L}^{2}({\cal R})\rightarrow{\cal L}^{2}({\cal
R}),\quad B_{q\mp}=\left[p+\pi_{q}\pm iW(q)\right],\quad
B_{p\mp}:{\cal L}^{2}({\cal R})\rightarrow{\cal L}^{2}({\cal
R}),\quad B_{p\mp}=\left[q+\pi_{p} \pm iV(p)\right].
\end{array}
\label {R226}
\end{equation}
In the expressions~(\ref{R226}), $W(q):{\cal R}\rightarrow{\cal R}$
and $V(p):{\cal R}\rightarrow{\cal R}$ are the piecewise
continuously differentiable functions called SUSY potentials. In
particular case if $R(q,\,p)=R_{q}(q)-R_{p}(p)$, for the evolution
of $B_{q,p\pm}$ we obtain
\begin{equation}
\begin{array}{l}
\dot{B}_{q\mp}=\left[-\left(F'_{q}+R'_{q}\,N\right)\pm
iW'_{q}(q)\left(p+\pi_{q}\right) \right],\quad
\dot{B}_{p\mp}=\left[-\left(G'_{p}+R'_{p}\,N\right)\pm
iV'_{p}(p)\left(q+\pi_{p}\right) \right].
\end{array}
\label {R228}
\end{equation}
Consequently,
\begin{equation}
\begin{array}{l}
\dot{Q}_{q\pm}=\pm i\left[\left(W'_{q}-R'_{q\pm}\right)\pm
i\left(F'_{q}-WW'_{q}\right)\, \psi_{\mp}\right],\quad
\dot{Q}_{p\pm}=\pm i\left[\left(V'_{p}-R'_{p\pm}\right)\pm
i\left(G'_{p}-V V'_{p}\right)\, \psi_{\mp}\right].
\end{array}
\label {R229}
\end{equation}
This shows that either $\dot{Q}_{q\pm}=0$ or $\dot{Q}_{p\pm}=0$ when
$W'_{q}(q)=R'_{q\pm}(q)$ and $F'_{q}=\frac{1}{2}(W^{2})'_{q}$ or
$V'_{p}(p)=R'_{p\pm}(p)$ and $G'_{p}=\frac{1}{2}(V^{2})'_{p}$,
respectively. Therefore, when the functions $R_{q,p}$ and
$F(q),\,G(p)$ are related as
\begin{equation}
\begin{array}{l}
R'_{q\pm}(q)=W'_{q}(q),\quad F_{q}=\frac{1}{2}(W^{2})+C_{q},\quad
R'_{p\pm}(p)=V'_{p}(p),\quad F_{p}=\frac{1}{2}(V^{2})+C_{p},
\end{array}
\label {R230}
\end{equation}
where $C_{q,p}$ are constants, then odd quantities $Q_{q,p\pm}$ are
integrals of motion in addition to $H_{ext}$ and $N.$ According to
Eq.~(\ref{R3}) and Eq.~(\ref{R322}), let us present $H_{ext}$ in the
form $H_{ext}=H_{q}-H_{p},$ where
\begin{equation}
\begin{array}{l}
H_{q}=\FFr{1}{2}\,\pi_{q}^{2}+F^{2}(q)+ R_{q}\,N,\quad
 H_{p}=\FFr{1}{2}\,\pi_{p}^{2}+G^{2}(p)+ R_{p}\,N.
\end{array}
\label {R232}
\end{equation}
Then, $Q_{q,p\pm}$ and $N$ together with the $H_{q}$ and $H_{p}$
form the classical analog of the extended phase space SUSY algebra
\begin{equation}
\begin{array}{l}
\left\{Q_{q,p+},\,Q_{q,p-}\right\}=-i (H_{q,p}-C_{q,p}),\quad
\left\{H_{q,p},\,Q_{q,p\pm}\right\}=\left\{Q_{q,p\pm},\,Q_{q,p\pm}\right\}=0,\\\\
\left\{N,\,Q_{q,p\pm}\right\}=\pm i Q_{q,p\pm},\quad
\left\{N,\,H_{q,p}\right\}=0,
\end{array}
\label {R233}
\end{equation}
with constants $C_{q,p}$ playing a role of a central charges in
$(q,p)-$ spaces, $N$ is classical analog of the grading operator.
Putting $C_{q}=C_{p}=0,$ we arrive at the classical analog of the
extended phase space SUSY quantum mechanics given by the extended
Lagrangian
\begin{equation}
\begin{array}{l}
{\cal L}_{ext}(p,\,q,\,\dot{p},\,\dot{q})=
\FFr{1}{2}\,\pi_{q}^{2}-\FFr{1}{2}\,W^{2}(q)+
\FFr{1}{2}\,\pi_{p}^{2}-\FFr{1}{2}\,V^{2}(p)+
\psi_{1}\psi_{2}(W'_{q}+V'_{p})+\FFr{1}{2}\psi_{\alpha}\dot{\psi}_{\alpha}.
\end{array}
\label{R3223}
\end{equation}
We conclude that the classical system Eq.~(\ref{R217}) is
characterized by the presence of two additional local in time odd
integrals of motion Eq.~(\ref{R225}) being supersymmetry generators.
Along the actual trajectories in $q-$space, the Eq.~(\ref{R3223})
reproduces the results obtained in~\cite{rf:4}.

\section{The path integral formulation} In the matrix formulation of extended
phase space (N=2)-SUSY quantum mechanics,  the $\hat{\psi}_{\pm}$
will be two real fermionic creation and annihilation nilpotent
operators describing the fermionic variables. The
$\hat{\psi}_{\pm}$, having anticommuting c-number eigenvalues, imply
\begin{equation}
\begin{array}{l}
\hat{\psi}_{\pm}=\sqrt{\FFr{1}{2}}\,\left(\hat{\psi}_{1}\pm
i\hat{\psi}_{2}\right),\quad
\left\{\hat{\psi}_{\alpha},\,\hat{\psi}_{\beta}\right\}=\delta_{\alpha\beta},\quad\left\{\hat{\psi}_{+},\,
\hat{\psi}_{-}\right\}=1,\quad \hat{\psi}_{\pm}^{2}=0.
\end{array}
\label {R103}
\end{equation}
They can be represented by finite dimensional matrices
$\hat{\psi}_{\pm}=\sigma^{\pm}$, where
$\sigma^{\pm}=\frac{\sigma_{1}\pm \sigma_{2}}{2}$ are the usual
raising and lowering operators for the eigenvalues of $\sigma_{3}$
which is the diagonal Pauli matrix. The fermionic operator $\hat{f}$
reads $\hat{f}:{\cal C}^{2}\rightarrow{\cal C}^{2},\quad
\hat{f}=\FFr{1}{2}\left[\hat{\psi}_{+},\,\hat{\psi}_{-}\right], $
which commutes with the $H_{ext}$ and is diagonal in this
representation with conserved eigenvalues $\pm\frac{1}{2}$. Due to
it the wave functions become two-component objects:
\begin{equation}
\begin{array}{l}
\chi(q,p)=
\left(
  \begin{array}{c}
   \chi_{+1/2}(q,p) \\
   \chi_{- 1/2}(q,p) \\
  \end{array}
\right)=\left(
          \begin{array}{c}
            \chi_{1}(q,p)\\
           \chi_{2}(q,p)  \\
          \end{array}
        \right)
= \left(
  \begin{array}{c}
   \psi_{1}(q)\phi_{1}(p)  \\
    \psi_{2}(q)\phi_{2}(p) \\
  \end{array}
\right),
\end{array}
\label {R387}
\end{equation}
where the states  $\psi_{1,2}(q),\,\phi_{1,2}(p)$ correspond to
fermionic quantum number $f=\pm \FFr{1}{2}$, respectively, in $q-$
and $p-$ spaces. They belong to Hilbert space ${\cal H}={\cal
H}_{0}\otimes{\cal C}^{2}=\left[{\cal L}^{2}({\cal R})\otimes{\cal
L}^{2}({\cal R})\right]\otimes{\cal C}^{2}. $ Hence the Hamiltonian
$H_{ext}$ of extended phase space (N=2)-SUSY quantum mechanical
system  becomes a $2\times2$ matrix:
\begin{equation}
\begin{array}{l}
H_{ext}=\left(
          \begin{array}{cc}
             H_{+} & 0 \\
            0 &  H_{-} \\
          \end{array}
        \right)=\FFr{1}{2}\left(\hat{\pi}_{q}^{2}+W^{2}(\hat{q})+
iW'_{q}(\hat{q})\left[\hat{\psi}_{1},\,\hat{\psi}_{2}\right]\right)-
\FFr{1}{2}\left(\hat{\pi}_{p}^{2}+V^{2}(\hat{p})+
iV'_{q}(\hat{p})\left[\hat{\psi}_{1},\,\hat{\psi}_{2}\right]\right).
\end{array}
\label {R208}
\end{equation}
To infer the extended Hamiltonian Eq.~(\ref{R208}) equivalently one
may start from the c-number extended Lagrangian of extended phase
space quantum field theory in $(0+1)$-dimensions in $q-$ and $p-$
spaces:
\begin{equation}
\begin{array}{l}
{\cal L}_{ext}(p,\,q,\,\dot{p},\,\dot{q})=-\dot{q}\,p -q\,\dot{p} +
\FFr{1}{2}\left[\left(\FFr{dq}{dt}\right)^{2}-W^{2}(q)\right]+f\,W'_{q}(q)+
\FFr{1}{2}\left[\left(\FFr{dp}{dt}\right)^{2}-V^{2}(p)\right]+f\,V'_{p}(p).
\end{array}
 \label{R104}
\end{equation}
In dealing with abstract space of eigenstates of the conjugate
operator $\hat{\psi}_{\pm}$ which have  anticommuting c-number
eigenvalues, suppose $|00->$ is the normalized zero-eigenstate of
$\hat{q}$ and $\hat{\psi}_{-}$:
\begin{equation}
\begin{array}{l}
\hat{q}|00->=0,\quad \hat{\psi}_{-}|00->=0.
\end{array}
\label {A1}
\end{equation}
The state $|00+>$ is defined by
\begin{equation}
\begin{array}{l}
|00+>=\hat{\psi}_{+}|00->=0,
\end{array}
\label {A2}
\end{equation}
then $\hat{\psi}_{+}|00+>=0,\quad \hat{\psi}_{-}|00+>=|00->. $
Taking into account that $\hat{\psi}_{\pm}^{\dag}=\hat{\psi}_{\mp},$
we get
\begin{equation}
\begin{array}{l}
<\mp00|\hat{\psi}_{\pm}=0,\quad <\mp00|\hat{\psi}_{\mp}=<\pm00|.
\end{array}
\label {A4}
\end{equation}
Now we may introduce the notation $\alpha, \beta,\dots$ for the
anticommuting eigenvalues of $\hat{\psi}_{\pm}.$ Consistency
requires:
\begin{equation}
\begin{array}{l}
\alpha\hat{\psi}_{\pm}=-\hat{\psi}_{\pm}\alpha,\quad
\alpha|00\pm>=\pm|00\pm> \alpha.
\end{array}
\label {A5}
\end{equation}
The eigenstates of $\hat{q}, \hat{\psi}_{-}$ can be constructed as
\begin{equation}
\begin{array}{l}
|q\alpha->=e^{-iq\hat{p}-\alpha\hat{\psi}_{+}}|00->,
\end{array}
\label {A6}
\end{equation}
and thus,
\begin{equation}
\begin{array}{l}
\hat{q}|q\alpha->=q |q\alpha->,\quad
\hat{\psi}_{-}|q\alpha->=\alpha|q\alpha->.
\end{array}
\label {A7}
\end{equation}
Then, the $\hat{\pi}_{q}$ and $ \hat{\psi}_{+}$ eigenstates  are
obtained by Fourier transformation:
\begin{equation}
\begin{array}{l}
|q\beta+>=-\IIn d\alpha\,e^{\alpha\beta}|q\alpha->,\quad
|\pi_{q}\alpha\pm>=-\IIn dq\,e^{iq\pi_{q}}|q\alpha\pm>,\\\\
|p\beta+>=-\IIn d\alpha\,e^{\alpha\beta}|p\alpha->,\quad
|\pi_{p}\alpha\pm>=-\IIn dp\,e^{ip\pi_{p}}|p\alpha\pm>,
\end{array}
\label {A8}
\end{equation}
which gives
\begin{equation}
\begin{array}{l}
\hat{\pi}_{q}|\pi_{q}\alpha\pm>=\pi_{q} |\pi_{q}\alpha\pm>,\quad
\hat{\psi}_{+}|(q,\pi_{q})\beta+>=\beta|(q,\pi_{q})\beta+>,\\\\
\hat{\pi}_{p}|\pi_{p}\alpha\pm>=\pi_{p} |\pi_{p}\alpha\pm>,\quad
\hat{\psi}_{+}|(p,\pi_{p})\beta+>=\beta|(p,\pi_{p})\beta+>.
\end{array}
\label {A9}
\end{equation}
The following completeness relations hold:
\begin{equation}
\begin{array}{l}
-\IIn d\alpha\,dq |q\alpha\pm><\mp \alpha^{*}q|=1,\quad -\IIn
d\alpha\,\FFr{d\pi_{q}}{2\pi} |\pi_{q}\alpha\pm><\mp
\alpha^{*}\pi_{q}|=1,\\\\
-\IIn d\alpha\,dp |p\alpha\pm><\mp \alpha^{*}p|=1,\quad -\IIn
d\alpha\,\FFr{d\pi_{p}}{2\pi} |\pi_{p}\alpha\pm><\mp
\alpha^{*}\pi_{p}|=1.
\end{array}
\label {A12}
\end{equation}
The time evolution of the state $|t>$ is now given
\begin{equation}
\begin{array}{l}
\chi_{-}(q\alpha\,p\beta\, t)=-\IIn\,d\alpha'\,dq'\,d\beta'\,dp'
K(q\alpha\,p\beta\, t|q'\alpha'\,p'\beta'\, t').
\end{array}
\label {R203}
\end{equation}
The kernel reads
\begin{equation}
\begin{array}{l}
K(q\alpha\,p\beta\, t|q'\alpha'\,p'\beta'\,
t')=<+q\alpha^{*}p\beta^{*}|e^{-iH_{ext}(t-t')}|q'\alpha'p'\beta'>,
\end{array}
\label {R204}
\end{equation}
which can be evaluated by the path integral. Actually, an
alternative approach to describe the state space and dynamics of the
extended phase space quantum system is by the path
integral~\cite{rf:2}, which reads
\begin{equation}
\begin{array}{l}
{\cal K}_{ff'}(qpt|q'p't')=<qpf|e^{-iH_{ext}(t-t')}|q'p'f'>.
\end{array}
\label {R184}
\end{equation}
In the path integral Eq.~(\ref{R184})  the individual states are
characterized by the energy and the fermionic quantum number $f.$
With the Hamiltonian $H_{ext}$, the path integral Eq.~(\ref{R184})
is diagonal:
\begin{equation}
\begin{array}{l}
{\cal K}_{ff'}(qpt|q'p't')= {\cal K}_{ff'}(qt|q't')\,{\cal
K}_{ff'}(pt|p't')=\delta_{ff'}\IIn_{q'}^{q}\,{\cal
D}q\,\IIn_{p'}^{p}\,{\cal D}p\,\exp{\left(i\IIn_{t'}^{t}\,{\cal
L}_{ext}(p,\,q,\,\dot{p},\,\dot{q})dt\right)}.
\end{array}
\label {R185}
\end{equation}
Knowing the path integral Eq.~(\ref{R185}), it is sufficient to
specify the initial wave function $\chi_{f}(q',p',t')$ to obtain all
possible information about the system at any later time $t,$ by
\begin{equation}
\begin{array}{l}
\chi_{f}(q,p,t)=\S_{f'}\IIn\,dq'\,dp'\,{\cal
K}_{ff'}(qpt|q'p't')\,\chi_{f'}(q',p',t'),
\end{array}
\label {R86}
\end{equation}
with $\chi_{\pm1/2}(q,p,t)=\chi_{1,2}(q,p,t)$ (Eq.~(\ref{R387})). In
terms of anticommuting c-number operators $\zeta$ and $\eta$
defining $\psi=\sqrt{\FFr{1}{2}}\left(
                                  \begin{array}{c}
                                    \eta+\zeta \\
                                    i(\eta-\zeta) \\
                                  \end{array}
                                \right), $ the path integral
Eq.~(\ref{R185}) becomes
\begin{equation}
\begin{array}{l}
{\cal K}(q\alpha p\beta t|q'\alpha' p'\beta' t')=
\IIn_{q',\alpha',p',\beta'}^{q,\alpha,p,\beta}\,{\cal D}q\,{\cal
D}p\,{\cal D}\zeta\,{\cal D}\eta\,\exp{\left(i\IIn_{t'}^{t}\,{\cal
L}_{ext}(p,\,q,\,\dot{p},\,\dot{q})dt\right)}.
\end{array}
\label {R219}
\end{equation}
The functional integral is taken over all trajectories from
$q',\alpha'$ to $q,\alpha$ and $p',\beta'$ to $p,\beta$ between the
times $t'$ and $t.$

\section{Solution of the extended Schr\"{o}dinger equation with
small energy eigenvalue $\varepsilon$ } Adopting the technique
developed in~\cite{rf:8}, first, we use  the iterative scheme to
find the approximate groundstate solutions to the extended
Schr\"{o}dinger-like equation
\begin{equation}
\begin{array}{l}
H_{ext}\,\chi(q,p)=(H_{q}-H_{p})\,\chi(q,p)=\varepsilon\,\chi(q,p),
\end{array}
\label {R59}
\end{equation}
with energy $\varepsilon$. We will then use these solutions to
calculate the parameters which measure the breaking of extended SUSY
such as the groundstate energy. The approximation, which went into
the derivation of solutions of Eq.~(\ref{R59}) meets our interest
that the groundstate energy $\varepsilon$ is supposedly small. As we
mentioned above the solutions for non-zero $\varepsilon$ come in
pairs of the form
\begin{equation}
\begin{array}{l}
\chi_{\uparrow}(q,p)=
\left(
  \begin{array}{c}
    \chi_{1}(q,p) \\
    0  \\
  \end{array}
\right)  \quad \mbox{or}\quad \chi_{\downarrow}(q,p)=
\left(
  \begin{array}{c}
    0 \\
    \chi_{2}(q,p) \\
  \end{array}
\right),
\end{array}
\label {R60}
\end{equation}
related by supersymmetry, where $\chi_{1,2}(q,p)=
\psi_{1,2}(q)\phi_{1,2}(p). $ The state space of the system is
defined by all the normalizable solutions of Eq.~(\ref{R59}) and the
individual states are characterized by the energies
$\varepsilon_{q}$ and $\varepsilon_{q}$ and the fermionic quantum
number $f$. One of these solutions is acceptable only if $W(q)$ and
$V(p)$ become infinite at both $q\rightarrow \pm\infty$ and
$p\rightarrow \pm\infty$, respectively, with the same sign. If this
condition is not satisfied, neither of the solutions is
normalizable, and they cannot represent the groundstate of the
system. The Eq.~(\ref{R59}) yields the following relations between
energy eigenstates with fermionic quantum number $\pm\frac{1}{2}$:
\begin{equation}
\begin{array}{l}
\left[\left(\FFr{\partial}{\partial q}+W_{q}(q)\right)-
\left(\FFr{\partial}{\partial
p}+V_{p}(p)\right)\right]\psi_{1}(q)\phi_{1}(p)=
\sqrt{2\varepsilon_{q}}\psi_{2}(q)\phi_{1}(p)-
\sqrt{2\varepsilon_{p}}\psi_{1}(q)\phi_{2}(p),
\end{array}
\label {R62}
\end{equation}
and
\begin{equation}
\begin{array}{l}
\left[\left(-\FFr{\partial}{\partial q}+W_{q}(q)\right)-
\left(-\FFr{\partial}{\partial
p}+V_{p}(p)\right)\right]\psi_{2}(q)\phi_{2}(p)=
\sqrt{2\varepsilon_{q}}\,\psi_{1}(q)\phi_{2}(p)-
\sqrt{2\varepsilon_{p}}\,\psi_{2}(q)\phi_{1}(p),
\end{array}
\label {R63}
\end{equation}
where $\varepsilon=\varepsilon_{q}-\varepsilon_{p},$
$\quad\varepsilon_{q}$ and $\varepsilon_{p}$ are the eigenvalues of
$H_{q}$ and $H_{p},$ respectively. The technique now is to devise an
iterative approximation scheme to solve Eq.~(\ref{R62}) and
Eq.~(\ref{R63}) by taking a trial wave function for $\chi_{2}(q,p),$
substitute this into the first equation ~(\ref{R62}) and integrate
it to obtain an approximation for $\chi_{1}(q,p).$ This can be used
as an ansatz in the second equation Eq.~(\ref{R63}) to find an
improved solution for $\chi_{2}(q,p),$ etc. As it was shown
in~\cite{rf:8}, the procedure converges for well-behaved potentials
with a judicious choice of initial trial function. If the $W_{q}$
and $V_{p}$ are odd, then
\begin{equation}
\begin{array}{l}
\psi_{1}(-q)=\psi_{2}(q), \quad \phi_{1}(-p)=\phi_{2}(p),
\end{array}
\label {R64}
\end{equation}
since they satisfy the same eigenvalue equation. It is
straightforward then, for example, to obtain
\begin{equation}
\begin{array}{l}
\left[\left(\FFr{\partial}{\partial q}+W_{q}(q)\right)-
\left(\FFr{\partial}{\partial
p}+V_{p}(p)\right)\right]\psi_{1}(q)\phi_{1}(p)=
\sqrt{2\varepsilon_{q}}\,\psi_{1}(-q)\phi_{1}(p)-
\sqrt{2\varepsilon_{p}}\,\psi_{1}(q)\phi_{1}(-p).
\end{array}
\label {R65}
\end{equation}
The independent nature of $q$ and $p$ gives the freedom of taking
$q=0,\quad p=0$ which yield an expression for energies:
\begin{equation}
\begin{array}{l}
\sqrt{2\varepsilon_{q}}=W(0)+\psi'_{1}(0)\left/\psi_{1}(0)\right.,
\quad
\sqrt{2\varepsilon_{p}}=V(0)+\phi'_{1}(0)\left/\phi_{1}(0)\right..
\end{array}
\label {R66}
\end{equation}
Suppose the potentials $W_{q}(q)$ and $V_{p}(p)$ have a maximum, at
$q_{-}$ and $p_{-}$, and minimum, at $q_{+}$ and $p_{+}$,
respectively. For the simplicity sake we choose the trial wave
functions as
\begin{equation}
\begin{array}{l}
\psi_{1,2}^{(0)}(q)=\delta(q-q_{\pm}),\quad
\phi_{1,2}^{(0)}(p)=\delta(p-p_{\pm}).
\end{array}
\label {R67}
\end{equation}
After one iteration, we obtain
\begin{equation}
\begin{array}{l}
\chi_{1}^{(1)}(q,p)=\FFr{1}{N_{q}N_{p}}\theta(q-q_{-})\,\theta(p-p_{-})\,e^{-\int_{0}^{q}\,dq'\,W(q')+
\int_{0}^{p}\,dp'\,V(p')},\\\\
\chi_{2}^{(1)}(q,p)=\FFr{1}{N_{q}N_{p}}\theta(q_{+}-q)\,\theta(p_{+}-p)\,e^{\int_{0}^{q}\,dq'\,W(q')-
\int_{0}^{p}\,dp'\,V(p')},
\end{array}
\label {R68}
\end{equation}
where $N_{q}$ and $N_{p}$ are the normalization factors. The next
approximation leads to
\begin{equation}
\begin{array}{l}
\chi_{1}^{(2)}(q,p)=\FFr{1}{N'}\,e^{-\int_{0}^{q}\,dq'\,W(q')+
\int_{0}^{p}\,dp'\,V(p')}\IIn_{max(-q,
q_{-})}^{\infty}e^{-2\int_{0}^{q'}\,dq''\,W(q'')}dq'\IIn_{max(-p,
p_{-})}^{\infty}e^{2\int_{0}^{p'}\,dp''\,V(p'')}dp',\\\\
\chi_{2}^{(2)}(q,p)=\FFr{1}{N'}\,e^{\int_{0}^{q}\,dq'\,W(q')-
\int_{0}^{p}\,dp'\,V(p')}\IIn^{min(-q,
q_{+})}_{\infty}e^{2\int_{0}^{q'}\,dq''\,W(q'')}dq'\IIn^{min(-p,
p_{+})}_{\infty}e^{-2\int_{0}^{p'}\,dp''\,V(p'')}dp'.
\end{array}
\label {R69}
\end{equation}
It can be easily verified that to this level of precision
Eq.~(\ref{R68}) is self-consistent solution. Actually, for example,
for $q'>q_{-}$ the exponential $e^{-2\int^{q'}_{0}W(q'')dq''}$  will
peak sharply around $q_{+}$ and may be approximated by a $\delta-$
function $c\delta(q_{+}-q');$ similarly
$e^{2\int^{q'}_{0}W(q'')dq''}$ we may replace approximately by
$c\delta(q_{-}-q')$ for $q'<q_{+}.$  The same arguments hold for the
p-space. With these approximations equations~(\ref{R69}) reduce to
Eq.~(\ref{R68}). The normalization constant $N'$ is
\begin{equation}
\begin{array}{l}
N'=\left(\IIn_{q_{-}}^{\infty}\,dq\,e^{-2\int_{0}^{q}W(q')dq'}\,
\IIn_{p_{-}}^{\infty}\,dp\,e^{-2\int_{0}^{p'}V(p')dp'}\right)^{3/2}.
\end{array}
\label {R70}
\end{equation}
The energy expectation value
\begin{equation}
\begin{array}{l}
\varepsilon=(\chi_{1},\,H_{ext}\,\chi_{1})
\end{array}
\label {R71}
\end{equation}
gives the same result as that obtained for odd potentials by means
of equations~(\ref{R66}) and~(\ref{R69}). Assuming the exponentials
$e^{-2\int_{q_{-}}^{0}W(q)dq}$ and $e^{-2\int_{p_{-}}^{0}V(p)dp}$ to
be small, which is correct to the same approximations underlying
Eq.~(\ref{R71}), the difference is negligible and the integrals in
both cases may be replaced by gaussians around $q_{+}$ and $p_{+}$,
respectively. Hence, it is straightforward to obtain
\begin{equation}
\begin{array}{l}
\varepsilon=\FFr{\hbar W'(q_{+})}{2\pi}\,e^{-2\Delta
W/\hbar}-\FFr{\hbar V'(p_{+})}{2\pi}\,e^{-2\Delta V/\hbar},
\end{array}
\label {R73}
\end{equation}
which gives direct evidence for the SUSY breaking in the extended
phase space quantum mechanical system. Here we have reinstated
$\hbar$, to show the order of adopted approximation, and its
non-perturbative nature. We also denoted
\begin{equation}
\begin{array}{l}
\Delta W=\IIn_{q_{+}}^{q_{-}}\,W(q)\, dq,\quad \Delta
V=\IIn_{p_{+}}^{p_{-}}\,V(p)\, dp.
\end{array}
\label {R74}
\end{equation}
However, a more practical measure for the SUSY breaking, in
particular, in field theories is the expectation value of an
auxiliary field, which can be replaced by its equation of motion
right from the start:
\begin{equation}
\begin{array}{l}
<F>=\left(\chi_{\uparrow},\,
i\{Q_{+},\,\sigma_{-}\}\chi_{\uparrow}\right).
\end{array}
\label {R75}
\end{equation}
Taking into account the relation $Q_{+}\chi_{\uparrow}=0, $ with
$Q_{+}$ commuting with $H_{ext}$, which means that the intermediate
state must have the same energy as $\chi$,  the Eq.~(\ref{R75}) can
be written in terms of a complete set of states as
\begin{equation}
\begin{array}{l}
<F>=i\left(\chi_{\uparrow},\,
Q_{+}\,\chi_{\downarrow}\right)\left(\chi_{\downarrow},\,
\sigma_{-},\,\chi_{\uparrow}\right).
\end{array}
\label {R77}
\end{equation}
According to Eq.~(\ref{R71}) we have
\begin{equation}
\begin{array}{l}
\varepsilon=<H_{ext}>=(\chi_{\uparrow},\,H_{ext}\,\chi_{\uparrow})=\FFr{1}{2}\left(\chi_{\uparrow},\,
Q_{+}\,\chi_{\downarrow}\right)\left(\chi_{\downarrow},\,
Q_{-}\,\chi_{\uparrow}\right)=\varepsilon_{q}-\varepsilon_{p}=
<H_{q}>-<H_{p}>=\\\\
(\psi_{\uparrow},\,H_{q}\,\psi_{\uparrow})-(\phi_{\uparrow},\,H_{p}\,\phi_{\uparrow})=
\FFr{1}{2}\left(\psi_{\uparrow},\,
Q_{q+}\,\psi_{\downarrow}\right)\left(\psi_{\downarrow},\,
Q_{q-}\,\psi_{\uparrow}\right)- \FFr{1}{2}\left(\phi_{\uparrow},\,
Q_{p+}\,\phi_{\downarrow}\right)\left(\phi_{\downarrow},\,
Q_{p-}\,\phi_{\uparrow}\right),
\end{array}
\label {R78}
\end{equation}
where
$\chi_{\uparrow\downarrow}=\psi_{\uparrow\downarrow}\phi_{\uparrow\downarrow},$
and $\psi_{\uparrow}=
\left(
  \begin{array}{c}
    \psi_{1} \\
    0 \\
  \end{array}
\right),\quad \psi_{\downarrow}=\left(
                                  \begin{array}{c}
                                    0 \\
                                    \psi_{2} \\
                                  \end{array}
                                \right),\quad
\phi_{\uparrow}=\left(
                  \begin{array}{c}
                    \phi_{1} \\
                    0 \\
                  \end{array}
                \right),\quad \phi_{\downarrow}=\left(
                                                  \begin{array}{c}
                                                    0 \\
                                                    \phi_{2} \\
                                                  \end{array}
                                                \right)$.  From SUSY algebra it follows
immediately that
\begin{equation}
\begin{array}{l}
\FFr{1}{\sqrt{\varepsilon}}\,Q_{-}\,\chi_{\uparrow}=\chi_{\downarrow}=
\hat{\psi}_{-}\,\chi_{\uparrow},\quad
\FFr{1}{\sqrt{\varepsilon_{q}}}\,Q_{q-}\,\psi_{\uparrow}=\psi_{\downarrow}=
\hat{\psi}_{-}\,\psi_{\uparrow},\quad
\FFr{1}{\sqrt{\varepsilon_{p}}}\,Q_{p-}\,\phi_{\uparrow}=\phi_{\downarrow}=
\hat{\psi}_{-}\,\phi_{\uparrow}.
\end{array}
\label {R80}
\end{equation}
By virtue of Eq.~(\ref{R80}), the Eq.~(\ref{R78}) reads
\begin{equation}
\begin{array}{l}
\sqrt{\varepsilon}\,\left(\chi_{\uparrow},\,
Q_{+}\,\chi_{\downarrow}\right)\left(\chi_{\downarrow},\,
\hat{\psi}_{-}\,\chi_{\uparrow}\right)=
\sqrt{\varepsilon_{q}}\,\left(\psi_{\uparrow},\,
Q_{q+}\,\psi_{\downarrow}\right)\left(\psi_{\downarrow},\,
\hat{\psi}_{-}\,\psi_{\uparrow}\right)-
\sqrt{\varepsilon_{p}}\,\left(\phi_{\uparrow},\,
Q_{p+}\,\phi_{\downarrow}\right)\left(\phi_{\downarrow},\,
\hat{\psi}_{-}\,\phi_{\uparrow}\right),
\end{array}
\label {R81}
\end{equation}
Using the matrix representations of $Q_{q+}, Q_{p+}$ and
$\hat{\psi}_{-}$ and the wave functions Eq.~(\ref{R69}), one
gets~(\cite{rf:8})
\begin{equation}
\begin{array}{l}
\left(\psi_{\downarrow},\,
\hat{\psi}_{-}\,\psi_{\uparrow}\right)=\sqrt{\FFr{W'(q_{+})}{\pi}}\,e^{-\Delta
W}\Delta q,\quad \left(\phi_{\downarrow},\,
\hat{\psi}_{-}\,\phi_{\uparrow}\right)=\sqrt{\FFr{V'(p_{+})}{\pi}}\,e^{-\Delta
V}\Delta p,\\\\
\left(\psi_{\uparrow},\, Q_{q+}\,\psi_{\downarrow}\right)=
i\sqrt{\FFr{W'(q_{-})}{\pi}}\,e^{-\Delta W},\quad
\left(\phi_{\uparrow},\, Q_{p+}\,\phi_{\downarrow}\right)=
i\sqrt{\FFr{V'(p_{-})}{\pi}}\,e^{-\Delta V},
\end{array}
\label {R82}
\end{equation}
where $\Delta q=q_{+}-q_{-}$ and $\Delta p=p_{+}-p_{-}.$ Hence
\begin{equation}
\begin{array}{l}
\left(\chi_{\uparrow},\,
Q_{+}\,\chi_{\downarrow}\right)\left(\chi_{\downarrow},\,
\hat{\psi}_{-}\,\chi_{\uparrow}\right)=2i\sqrt{\varepsilon}\,
\left[\left(\FFr{\varepsilon_{q}}{\varepsilon}\right)\,
\sqrt{\varepsilon}_{q}\,\sqrt{\FFr{W'(q_{-})}{\pi}}\,\Delta\,q-
\left(\FFr{\varepsilon_{p}}{\varepsilon}\right)\,
\sqrt{\varepsilon}_{p}\,\sqrt{\FFr{V'(p_{-})}{\pi}}\,\Delta\, p
\right],
\end{array}
\label {R83}
\end{equation}
and
\begin{equation}
\begin{array}{l}
<F>=
-2\sqrt{\varepsilon}\,\left[\left(\FFr{\varepsilon_{q}}{\varepsilon}\right)\,
\sqrt{\FFr{W'(q_{+})}{\pi}}\,e^{-2\Delta W}\Delta q -
\left(\FFr{\varepsilon_{p}}{\varepsilon}\right)\,
\sqrt{\FFr{V'(p)}{\pi}}\,e^{-2\Delta V}\Delta p\right].
\end{array}
\label {R84}
\end{equation}
Along the actual trajectories in $q-$space, the Eq.~(\ref{R84})
reproduces the results obtained in~\cite{rf:8}.

\section{An extended SUSY breaking in the instanton picture} In this subsection
our goal is to show that the expressions Eq.~(\ref{R83}) and
Eq.~(\ref{R84}) can be obtained in the path integral formulation of
the theory by calculating the matrix elements, i.e., the effect of
tunneling between two classical vacua by using a one-instanton
background. That is, the matrix elements of $\hat{\psi}_{\pm},
Q_{q\pm}$ and $Q_{p\pm}$ can be calculated in the background of the
classical solution $\dot{q}_{c}=-W_{c}$ and $\dot{p}_{c}=-V_{c}$. In
doing this we re-write the matrix element Eq.~(\ref{R81}) in terms
of eigenstates of the conjugate operator $\hat{\psi}_{\pm}$:
\begin{equation}
\begin{array}{l}
\sqrt{\varepsilon}\,\left[<+0q_{+}p_{+}|\,e^{-iH_{ext}(T-t)}\,Q_{+}e^{-iH_{ext}(T+t)}\,|q_{-}p_{-}0->
<-0q_{-}p_{-}|\,e^{-iH_{ext}(T-t)}\,\hat{\psi}_{-}\,e^{-iH_{ext}(T+t)}\,|q_{+}p_{+}0+>\right]_{T\rightarrow-i\infty}=
\\\\
\sqrt{\varepsilon_{q}}\,\left[
<+0q_{+}|\,e^{-iH_{q}(T-t)}\,Q_{q+}\,e^{-iH_{q}(T+t)}\,|q_{-}0->
<-0q_{-}|\,e^{-iH_{q}(T-t)}\,\hat{\psi}_{-}|\,e^{-iH_{q}(T+t)}\,q_{+}0+>\right]_{T\rightarrow-i\infty}-\\\\
\sqrt{\varepsilon_{p}}\,\left[
<+0p_{+}|\,e^{-iH_{p}(T-t)}\,Q_{p+}\,e^{-iH_{p}(T+t)}\,|p_{-}0->
<-0p_{-}|\hat{\psi}_{-}|p_{+}0+>\right]_{T\rightarrow-i\infty},
\end{array}
\label {R85}
\end{equation}
in the limit $T\rightarrow -i\infty$. This reduces to
\begin{equation}
\begin{array}{l}
\sqrt{\varepsilon}\,<+0q_{+}p_{+}|Q_{+}|q_{-}p_{-}0->
<-0q_{-}p_{-}|\hat{\psi}_{-}|q_{+}p_{+}0+>= \sqrt{\varepsilon_{q}}\,
<+0q_{+}|Q_{q+}|q_{-}0->
<-0q_{-}|\hat{\psi}_{-}|q_{+}0+>-\\\\
\sqrt{\varepsilon_{p}}\,<+0p_{+}|Q_{p+}|p_{-}0->
<-0p_{-}|\hat{\psi}_{-}|p_{+}0+>,
\end{array}
\label {R86}
\end{equation}
which, in turn, can be presented by path integrals defined in terms
of anticommuting c-number operators $\zeta$ and $\eta$ with
Euclidean actions of the instantons in $q-$ and $p-$ spaces,
respectively. Following~\cite{rf:8}, these functional integrals
include an integration over {\it instanton time} $\tau_{0}$ which is
due to the problem of zero modes of the bilinear terms in Euclidean
actions. This arises from time-transformation of instantons, and
SUSY transformations on them, respectively. The existence of zero
modes gives rise to non-gaussian behaviour of the functional
integral. Due to it the matrix elements above do not receive any
contributions from either no-instanton or anti-instanton
configurations. The zero mode problem is solved by introducing a
collective coordinate $\tau_{0}$ replacing the bosonic zero
mode~\cite{rf:9}. Whereas, the funcional integrals  depend only on
the difference $\tau-\tau_{0}.$  Note also that multi-instanton
configurations could contribute in principle, provided they have not
more than one normalizable fermionic zero mode. But as it was shown
in~\cite{rf:8}, their contribution is clearly smaller with respect
to $\sqrt{2\varepsilon_{q}}$ and $\sqrt{2\varepsilon_{p}}$. In the
case when the SUSY potentials in $q-$ and $p-$ spaces have more than
two extrema $q_{\nu}$ and $p_{\mu}$, $\nu, \mu=1,2,\dots, N$, one
can put conditions on the SUSY potentials
\begin{equation}
\begin{array}{l}
\IIn_{0}^{\infty}W(q')dq' \rightarrow \infty \quad\mbox{at}\quad
q\rightarrow \pm \infty \quad\mbox{for}\quad \psi_{0}^{+},\quad
\IIn_{0}^{\infty}W(q')dq' \rightarrow -\infty \quad\mbox{at}\quad
q\rightarrow \pm \infty \quad\mbox{for}\quad \psi_{0}^{-},
\end{array}
\label {R42}
\end{equation}
and similar for $V(p)$, that the extrema are well separated:
$\int_{q_{\nu}}^{q_{\nu}+1}W(q')dq'\gg 1$, and
$\int_{p_{\mu}}^{q_{\mu}+1}V(p')dp'\gg 1$. Around each of the
classical minima $q_{\nu}$ and $p_{\mu}$ of the potentials
$W^{2}(q)$ and $V^{2}(p)$, respectively, one can approximate the
theory by a suppersymmetric harmonic oscillator. Then there are $N$
ground states which have zero energy. These states are described by
upper or lower component of the wave function, depending on whether
$\nu$ and $\mu$ are odd or even. With this provision the functional
integrals are calculated in~\cite{rf:8}, which allow us consequently
to write:
\begin{equation}
\begin{array}{l}
<+0q_{+}|Q_{q+}|q_{-}0->=i\sqrt{\FFr{W'_{c}(q_{+})}{\pi}}\,e^{-\Delta
W_{c}},\quad
<-0q_{-}|\hat{\psi}_{-}|q_{+}0+>=\sqrt{\FFr{W'_{c}(q_{+})}{\pi}}\,e^{-\Delta
W_{c}}\Delta q_{c},
\end{array}
\label {R96}
\end{equation}
etc. Inserting this in Eq.~(\ref{R86}), we arrive at the
Eq.~(\ref{R83})
\begin{equation}
\begin{array}{l}
<+0q_{+}p_{+}|Q_{+}|q_{-}p_{-}0->
<-0q_{-}p_{-}|\hat{\psi}_{-}|q_{+}p_{+}0+>=
\\\\
2i\sqrt{\varepsilon}\,\left[\left(\FFr{\varepsilon_{q}}{\varepsilon}\right)\,
\sqrt{\varepsilon_{q}}\,\sqrt{\FFr{W'_{c}(q_{+})}{\pi}}\,\Delta
q_{c}- \left(\FFr{\varepsilon_{p}}{\varepsilon}\right)\,
\sqrt{\varepsilon_{p}}\,\sqrt{\FFr{V'_{c}(p_{+})}{\pi}}\,\Delta
p_{c}\right],
\end{array}
\label {R866}
\end{equation}
and, thus,  of Eq.~(\ref{R84}) as its inevitable corollary. This
proves that the extended SUSY breaking has resulted from tunneling
between the classical vacua of the theory. The corrections to this
picture are due to higher order terms and quantum tunneling effects.

\section{Spectrum generating algebra}
An extended Hamiltonian $H_{ext}$ Eq.~(\ref{R208}) can be treated as
a set of two ordinary two-dimensional partner
Hamiltonians~\cite{rf:10}
\begin{equation}
\begin{array}{l}
H_{\pm}=\FFr{1}{2}\left[\pi_{q}^{2}-\pi_{p}^{2}+
U_{\pm}(q,p)\right],
\end{array}
\label {R2}
\end{equation}
provided by partner potentials
\begin{equation}
\begin{array}{l}
U_{\pm}(q,p)=U_{q\pm}(q)-U_{p\pm}(p),\quad
U_{q\pm}(q)=W^{2}(q,\,a_{q})\mp W'_{q}(q,\,a_{q}),\quad
U_{p\pm}(p)=V^{2}(p,\,a_{p})\mp V'_{p}(p,\,a_{p}).
\end{array}
\label {R333}
\end{equation}
A subset of the SUSY potentials for which the Schr\"{o}dinger-like
equations are exactly solvable share an integrability conditions of
{\it shape-invariance}~\cite{rf:100}:
\begin{equation}
\begin{array}{l}
U_{+}(a_{0},q,p)=U_{-}(a_{1},q,p)+R(a_{0}), \quad a_{1}=f(a_{0}),
\end{array}
\label {R444}
\end{equation}
where $a_{0}$ and $a_{1}$ are a set of parameters that specify
phase-space-independent properties of the potentials, and the
reminder $R(a_{0})$ is independent of $(q,p).$

\subsection{Algebraic properties of shape invariance}
Using the standard technique, we may construct a series of
Hamiltonians $H_{N}, \quad N=0,1,2,\dots,$
\begin{equation}
\begin{array}{l}
H_{N}=\FFr{1}{2}\left[\pi_{q}^{2}-\pi_{p}^{2}+
U_{-}(a_{N},q,p)+\S_{k=1}^{N}R(a_{k})\right],
\end{array}
\label {R555}
\end{equation}
where $a_{N}=f^{(N)}(a)$ ($N$ is the number of iterations). From
Eqs.~(\ref{R333}) and (\ref{R444}) we obtain then $N=n+m$ coupled
nonlinear differential equations which are the two recurrence
relations of Riccati-type differential equations:
\begin{equation}
\begin{array}{l}
W^{2}_{n+1}(q)+W'_{q(n+1)}(q)=W^{2}_{n}(q)-W'_{qn}(q)-\mu_{n},\quad
V^{2}_{m+1}(p)+V'_{p(m+1)}(p)=V^{2}_{m}(p)-V'_{pm}(p)-\nu_{m},
\end{array}
\label {R666}
\end{equation}
where we denote $W_{n}(q)\equiv W(q,\,a_{qn}), \quad V_{m}(p)\equiv
V(p,\,a_{pm}),$ $\mu_{n}\equiv R_{q}(a_{qn})$ and $\nu_{m}\equiv
R_{p}(a_{pm}).$ Here we admit that for unbroken SUSY, the
eigenstates of the potentials $U_{q,p},$ respectively,  are
\begin{equation}
\begin{array}{l}
E^{(-)}_{q0}=0,\quad E^{(-)}_{qn}=\S_{i=0}^{n-1}\mu_{i},\quad
E^{(-)}_{p0}=0,\quad E^{(-)}_{pm}=\S_{j=0}^{m-1}\nu_{j},
\end{array}
\label {R7}
\end{equation}
that is, the ground states are at zero energies, characteristic of
unbroken supersymmetry. The differential equations~(\ref{R666}) can
be investigated to find exactly solvable potentials. The shape
invariance condition Eq.~(\ref{R444}) can be expressed in terms of
bosonic operators as
\begin{equation}
\begin{array}{l}
B_{+}(x,\,a_{0})B_{-}(x,\,a_{0})-B_{-}(x,\,a_{1})B_{+}(x,\,a_{1})=R(a_{0})=\mu_{0}-\nu_{0},\\\\
B_{{\ae}+}({\ae},\,a_{{\ae}0})B_{{\ae}-}({\ae},\,a_{{\ae}0})-
B_{{\ae}-}({\ae},\,a_{{\ae}1})B_{{\ae}+}({\ae},\,a_{{\ae}1})=(\mu_{0}\,\mbox{or}\,\nu_{0}),
\end{array}
\label {R10}
\end{equation}
where $x(q,p)$-is the coordinate in $(q,p)-$ space, ${\ae}$ denotes
concisely either $q-$ or $p-$ representations (no summation on
${\ae}$ is assumed).  To classify algebras associated with the shape
invariance, following~\cite{rf:12} we introduce an auxiliary
variables $\phi(\phi_{q},\,\phi_{p})$ and define the following
creation and annihilation operators:
\begin{equation}
\begin{array}{l}
J_{+}=e^{ik\phi}B_{+}(x,\,\chi(i\partial_{\phi})),\quad
J_{-}=B_{-}(x,\,\chi(i\partial_{\phi}))e^{-ik\phi},
\end{array}
\label {R11}
\end{equation}
where $k(k_{q},\,k_{p})$ are an arbitrary real constants and
$\chi(\chi_{q},\,\chi_{p})$ are an arbitrary real functions.
Consequently, the creation and annihilation operators in
$(q,p)-$spaces can be written as
\begin{equation}
\begin{array}{l}
J_{{\ae}+}=e^{ik_{{\ae}}\phi_{{\ae}}}B_{{\ae}+}({\ae},\,\chi_{{\ae}}(i\partial_{\phi_{{\ae}}})),\quad
J_{{\ae}-}=B_{{\ae}-}({\ae},\,\chi_{{\ae}}(i\partial_{\phi_{{\ae}}}))e^{-ik_{{\ae}}\phi_{{\ae}}}.
\end{array}
\label {R12}
\end{equation}
The operators $B_{\pm}(x,\,\chi(i\partial_{\phi}))$ and
$B_{{\ae}\pm}({\ae},\,\chi_{{\ae}}(i\partial_{\phi_{{\ae}}}))$ are
the generalization of Eqs.~(\ref{R10}), where $a_{0}\rightarrow
\chi(i\partial_{\phi})$ and $a_{{\ae}0}\rightarrow
\chi_{{\ae}}(i\partial_{\phi_{{\ae}}}).$  One can easily prove the
following relations:
\begin{equation}
\begin{array}{l}
e^{ik\phi}B_{+}(x,\,\chi(i\partial_{\phi}))=B_{+}(x,\,\chi(i\partial_{\phi}+k))e^{ik\phi},\quad
B_{-}(x,\,\chi(i\partial_{\phi}))e^{-ik\phi}=e^{-ik\phi}B_{-}(x,\,\chi(i\partial_{\phi}+k)),
\end{array}
\label {R13}
\end{equation}
and that
\begin{equation}
\begin{array}{l}
e^{ik_{{\ae}}\phi_{{\ae}}}B_{{\ae}+}({\ae},\,\chi_{{\ae}}(i\partial_{\phi_{{\ae}}}))=
B_{{\ae}+}({\ae},\,\chi_{{\ae}}(i\partial_{\phi_{{\ae}}}+k_{{\ae}}))e^{ik_{{\ae}}\phi_{{\ae}}},\\\\
B_{{\ae}-}({\ae},\,\chi_{{\ae}}(i\partial_{\phi_{{\ae}}}))e^{-ik_{{\ae}}\phi_{{\ae}}}=
e^{-ik_{{\ae}}\phi_{{\ae}}}B_{{\ae}-}({\ae},\,\chi_{{\ae}}(i\partial_{\phi_{{\ae}}}+k_{{\ae}})).
\end{array}
\label {R14}
\end{equation}
If we choose a function $\chi(i\partial_{\phi})$ such that
$\chi(i\partial_{\phi}+k)=f[\chi(i\partial_{\phi})],$ then we have
identified $a_{0}\rightarrow \chi(i\partial_{\phi}),$ and
$a_{1}=f(a_{0})\rightarrow
f[\chi(i\partial_{\phi})]=\chi(i\partial_{\phi}+k).$ Similar
relations can be obtained for the ${\ae}-$representations. From
Eq.~(\ref{R10}) we obtain then
\begin{equation}
\begin{array}{l}
B_{+}(x,\,\chi(i\partial_{\phi}))B_{-}(x,\,\chi(i\partial_{\phi}))-
B_{-}(x,\,\chi(i\partial_{\phi}+k))B_{+}(x,\,\chi(i\partial_{\phi}+k))=R[\chi(i\partial_{\phi}))],\\\\
B_{{\ae}+}({\ae},\,\chi_{{\ae}}(i\partial_{\phi_{{\ae}}}))B_{{\ae}-}({\ae},\,\chi(i\partial_{\phi_{{\ae}}}))-
B_{{\ae}-}({\ae},\,\chi_{{\ae}}(i\partial_{\phi_{{\ae}}}+k_{{\ae}}))
B_{{\ae}+}({\ae},\,\chi_{{\ae}}(i\partial_{\phi_{{\ae}}}+k_{{\ae}}))=R_{{\ae}}[\chi_{{\ae}}(i\partial_{\phi_{{\ae}}}))].
\end{array}
\label {R15}
\end{equation}
Introducing the operators $J_{3}=-\FFr{i}{k}\partial_{\phi}$ and
$J_{{\ae}3}=-\FFr{i}{k_{{\ae}}}\partial_{\phi_{{\ae}}}$, and
combining Eqs.~(\ref{R12}) and~(\ref{R15}), we may arrive at a
deformed Lie algebras:
\begin{equation}
\begin{array}{l}
\left[J_{+},\,J_{-}\right]=\left[J_{q+},\,J_{q-}\right]-\left[J_{p+},\,J_{p-}\right]=
\xi(J_{3})=\xi_{q}(J_{q3})-\xi_{p}(J_{p3}),\\\\
\left[J_{3},\,J_{\pm}\right]=\pm J_{\pm},\quad
\left[J_{{\ae}+},\,J_{{\ae}-}\right]=\xi_{{\ae}}(J_{{\ae}3}),\quad
\left[J_{{\ae}3},\,J_{{\ae}\pm}\right]=\pm J_{{\ae}\pm},
\end{array}
\label {R16}
\end{equation}
where $\xi(J_{3})\equiv R[\chi(i\partial_{\phi})]$ and
$\xi_{{\ae}}(J_{{\ae}3})\equiv
R_{{\ae}}[\chi_{{\ae}}(i\partial_{\phi_{{\ae}}})]$ define the
deformations. Different $\chi$ functions in Eq.~(\ref{R15}) define
different reparametrizations corresponding to several
models. For example: \\
1. The translational models ($a_{1}=a_{0}+k$) correspond to
$\chi(z)=z.$ If $R$ is a linear function of $J_{3}$ the algebra
becomes SO(2.1) or SO(3). Similar in many respects prediction is
made in
somewhat different method by Balantekin~\cite{rf:13}.\\
2. The scaling models ($a_{1}=e^{k}a_{0}$) correspond to
$\chi(z)=e^{z},$  etc.

\subsection{The unitary representations of the deformed Lie algebra}
In order to find the energy spectrum of the partner SUSY
Hamiltonians
\begin{equation}
\begin{array}{l}
2H_{-}(x,\,\chi(i\partial_{\phi}))=B_{-}(x,\,\chi(i\partial_{\phi}+k))B_{+}(x,\,\chi(i\partial_{\phi}+k)),\\\\
2H_{{\ae}-}({\ae},\,\chi_{{\ae}}(i\partial_{\phi_{{\ae}}}))
=B_{{\ae}-}({\ae},\,\chi_{{\ae}}(i\partial_{\phi_{{\ae}}}+k_{{\ae}}))
B_{{\ae}+}({\ae},\,\chi_{{\ae}}(i\partial_{\phi_{{\ae}}}+k_{{\ae}})),
\end{array}
\label {R17}
\end{equation}
one must construct the unitary representations of deformed Lie
algebra defined by Eq.~(\ref{R16})~\cite{rf:12,rf:111}. Using the
standard technique, one defines up to additive constants the
functions $g(J_{3})$ and $g_{{\ae}}(J_{{\ae}3})$:
\begin{equation}
\begin{array}{l}
\xi(J_{3})=g(J_{3})-g(J_{3}-1), \quad
\xi_{{\ae}}(J_{{\ae}3})=g_{{\ae}}(J_{{\ae}3})-g_{{\ae}}(J_{{\ae}3}-1).
\end{array}
\label {R18}
\end{equation}
The Casimirs of this algebra can be written as
$C_{2}=J_{-}J_{+}+g(J_{3})$ and accordingly
$C_{{\ae}2}=J_{{\ae}-}J_{{\ae}+}+g_{{\ae}}(J_{{\ae}3}).$ In a basis
in which $J_{3}$ and $C_{2}$ are diagonal, $J_{-}$ and $J_{+}$ are
lowering and raising operators (the same holds for
${\ae}$-representations). Operating on an arbitrary state $|h>$ they
yield
\begin{equation}
\begin{array}{l}
J_{3}|h>=h|h>,\quad  J_{-}|h>=a(h)|h-1>,\quad
J_{+}|h>=a^{*}(h+1)|h+1>,
\end{array}
\label {R19}
\end{equation}
where
\begin{equation}
\begin{array}{l}
|a(h)|^{2}-|a(h+1)|^{2}=g(h)-g(h-1).
\end{array}
\label {R20}
\end{equation}
Similar arguments can be used for the operators $J_{(p,q)3}$,
$C_{{\ae}2}$ and $J_{{\ae}\pm}$, which yield  similar relations for
the states $h_{{\ae}}.$ The profile of $g(h)$ (and, thus, of
$g_{{\ae}}(h_{{\ae}})$) determines the dimension of the unitary
representation. Having the representation of the algebra associated
with a characteristic model, consequently we obtain the complete
spectrum of the system.  For example,  without ever referring to
underlying differential equation, we may obtain analytic expressions
for the entire energy spectrum of extended Hamiltonian with {\it
Self-Similar} potential. A scaling change of parameters is given as
$a_{1}=Qa_{0}, \quad a_{{\ae}1}=Q_{{\ae}}a_{{\ae}0},$ at the simple
choice $R(a_{0})=-r_{1}a_{0},$ where $r_{1}$ is a constant. That is,
\begin{equation}
\begin{array}{l}
\xi(J_{3})\equiv -r_{1}\exp
(-kJ_{3})=\xi_{q}(J_{q3})-\xi_{p}(J_{p3})= -r_{q1}\exp
(-k_{q}J_{q3})+r_{p1}\exp (-k_{p}J_{p3}),
\end{array}
\label {R21}
\end{equation}
which yields
\begin{equation}
\begin{array}{l}
\left[J_{+},\,J_{-}\right]= \xi(J_{3})=-r_{1}\exp(-kJ_{3}),\quad
\left[J_{3},\,J_{\pm}\right]=\pm J_{\pm}.
\end{array}
\label {R21}
\end{equation}
This is a deformation of the standard $SO(2.1)$ Lie algebra,
therefore, one gets
\begin{equation}
\begin{array}{l}
g(h)=\FFr{r_{1}}{e^{k}-1}e^{-kh}=-\FFr{r_{1}}{1-Q}Q^{-h},\quad
Q=e^{k},\\\\
g_{{\ae}}(h_{{\ae}})=\FFr{r_{{\ae}1}}{e^{k_{{\ae}}}-1}e^{-k_{{\ae}}h_{{\ae}}}=
-\FFr{r_{{\ae}1}}{1-Q_{{\ae}}}Q_{{\ae}}^{-h_{{\ae}}},\quad
Q_{{\ae}}=e^{k_{{\ae}}}.
\end{array}
\label {R22}
\end{equation}
For scaling problems~\cite{rf:12} one has $0<q<1,$ which leads to
$k<0$. The unitary representation of this algebra for monotonically
decreasing profile of the function $g(h)$, are infinite dimensional.
Let the lowest weight state of the $J_{3}$ be $h_{min},$ then
$a(h_{min})=0.$ One can choose the coefficients $a(h)$ to be real.
From Eq.~(\ref{R20}), for an arbitrary $h=h_{min}+n, n=0,1,2,\dots$,
we obtain
\begin{equation}
\begin{array}{l}
a(h)^{2}=g(h-n-1)-g(h-1)=r_{1}\FFr{Q^{n}-1}{Q-1}Q^{1-h}.
\end{array}
\label {R23}
\end{equation}
The spectrum of the extended Hamiltonian $H_{-}(x,a_{1})$ reads
\begin{equation}
\begin{array}{l}
H_{-}|h>=a(h)^{2}|h>=r_{1}\FFr{Q^{n}-1}{Q-1}Q^{1-h}|h>,
\end{array}
\label {R24}
\end{equation}
with the eigenenergies
\begin{equation}
\begin{array}{l}
E_{n}(h)=r_{1}\alpha(h)\FFr{Q^{n}-1}{Q-1},\quad \alpha(h)\equiv
Q^{1-h}.
\end{array}
\label {R25}
\end{equation}
Similar expressions can be obtained for the $H_{{\ae}-}$ and
eigenenergies $E_{qn_{1}}$ and $E_{pn_{2}}$ ($
n\equiv(n_{1},n_{2}),\quad n_{1,2}=0,1,2,\dots$),  as
\begin{equation}
\begin{array}{l}
H_{q-}|h_{q}>=a_{q}(h_{q})^{2}|h_{q}>=r_{q1}\FFr{Q_{q}^{n_{1}}-1}{Q_{q}-1}Q_{q}^{1-h_{q}}|h_{q}>,\\\\
H_{p-}|h_{p}>=a_{p}(h_{p})^{2}|h_{p}>=r_{p1}\FFr{Q_{p}^{n_{2}}-1}{Q_{p}-1}Q_{p}^{1-h_{p}}|h_{p}>,
\end{array}
\label {R26}
\end{equation}
and that
\begin{equation}
\begin{array}{l}
E_{qn_{1}}(h_{q})=r_{q1}\alpha_{q}(h_{q})\FFr{Q_{q}^{n_{1}}-1}{Q_{q}-1},\quad
\alpha_{q}(h_{q})\equiv Q_{q}^{1-h_{q}},\\\\
E_{pn_{2}}(h_{p})=r_{p1}\alpha_{p}(h_{p})\FFr{Q_{p}^{n_{2}}-1}{Q_{p}-1},\quad
\alpha_{p}(h_{p})\equiv Q_{p}^{1-h_{p}}.
\end{array}
\label {R27}
\end{equation}
Hence
\begin{equation}
\begin{array}{l}
E_{n}(h)=E_{qn_{1}}(h_{q})- E_{pn_{2}}(h_{p}).
\end{array}
\label {R28}
\end{equation}

\section{Conclusions}
We addressed the classical analog of the extended phase space
quantum mechanics of particle which have both bosonic and fermionic
degrees of freedom, i.e., the particle with odd degrees of freedom
which gives rise to (N=2)-realization of the supersymmetry algebra.
We obtain the integrals of motion.  We use the iterative scheme to
find the approximate groundstate solutions to the extended
Schr\"{o}dinger-like equation and calculate the parameters which
measure the breaking of extended SUSY such as the groundstate
energy. The approximation, which went into the derivation of
solutions of Eq.~(\ref{R59}) meets our interest that the groundstate
energy $\varepsilon$ is supposedly small. This gives direct evidence
for the SUSY breaking. However, we calculate a more practical
measure for the SUSY breaking, in particular in field theories which
is the expectation value of an auxiliary field. We analyze in detail
the non-perturbative mechanism for extended phase space SUSY
breaking in the instanton picture and show that this has resulted
from tunneling between the classical vacua of the theory. Finally,
we present an analysis on the independent group theoretical methods
with nonlinear extensions of lie algebras from the extended phase
space SUSY quantum mechanics. Using the factorization procedure we
explore the algebraic property of shape invariance and spectrum
generating algebra. Most of these Hamiltonians posses this feature
and hence are solvable by an independent group theoretical method.
We construct the unitary representations of the deformed Lie
algebra.

\section*{Acknowledgments}
I would like to thank Y. Sobouti for drawing my attention to the
extended phase space formulation of quantum mechanics.

\end{document}